\documentclass[conference]{IEEEtran}
\IEEEoverridecommandlockouts

\usepackage{graphicx, multicol, multirow}
\graphicspath{{_figures/}}
\usepackage{booktabs} 
\usepackage{hyperref}	
\hypersetup{colorlinks=true, citecolor=blue, linkcolor=blue, urlcolor=blue}
\usepackage{amssymb, amsmath}

\title{Decoding Semantic Categories from Picture-Naming EEG}
\author{
    \IEEEauthorblockN{
    Wei Hu\IEEEauthorrefmark{1},
    Binbin Xu\IEEEauthorrefmark{4}
    }
    \IEEEauthorblockA{\IEEEauthorrefmark{1}College of Science and Technology, University of Bordeaux, France}
    \IEEEauthorblockA{\IEEEauthorrefmark{4}EuroMov Digital Health in Motion, Univ Montpellier, IMT Mines Ales, France}
    \IEEEauthorblockA{
    \small
    \texttt{binbin.xu@mines-ales.fr}
    }
}

\begin{document}

\maketitle

\begin{abstract}
Picture naming requires the transformation of visual object information into a spoken lexical response through perceptual, semantic, lexical, and articulatory processes. This study asked whether semantic-category information is recoverable from high-density EEG during overt picture naming. Sixteen native French-speaking participants performed a picture-naming task using line drawings. Picture labels were embedded with a multilingual text-embedding model and organized into nine interpretable semantic categories, providing a data-driven semantic target space for neural decoding. EEG activity was represented channel-wise using a pre-trained single-channel EEG encoder over an early post-stimulus window, a later naming-related window, and their combination. Nine-class decoding showed above-chance semantic-category discrimination in all temporal representations. Balanced accuracy increased from $0.562$ in the early window to $0.610$ in the naming-related window, and reached $0.781$ when both windows were combined, with a maximum Macro-F1 of $0.784$. Class-level F1 scores showed consistent gains across semantic categories, and sensor-level decoding maps indicated spatially distributed category information. These findings suggest that semantic-category structure is reflected in EEG activity during overt picture naming and that early and naming-related temporal windows provide complementary information. The results support the use of modern neural decoding methods as tools for investigating lexical-semantic processing in spoken language production.
\end{abstract}

\section{Introduction}

Picture naming is a central paradigm for studying spoken language production because it links a visually presented object to a lexical response under controlled experimental timing. Naming a picture requires visual object processing, access to conceptual and semantic information, lexical selection, phonological encoding, articulatory preparation, and overt speech production \cite{Indefrey2004spatial,Indefrey2011spatial,Levelt1999Theory}. Electrophysiological studies have used this paradigm to characterize the temporal dynamics of spoken word production, including early object processing, lexical-semantic access, and later response-related stages \cite{Schmitt2000Electrophysiological,Volpert2023Characterization}. This temporal structure makes picture naming a useful test case for asking whether semantic information is detectable in non-invasive neural recordings during language production.

Decoding semantic information from EEG is challenging because the relevant neural activity is distributed across time and sensors, while the measured signal is noisy, non-stationary, and sensitive to participant, session, and artifact variability \cite{Urigueen2015EEG}. These limitations are especially important in overt naming, where semantic and lexical processes are followed by articulatory planning, speech execution, and speech-related muscular activity \cite{Ganushchak2011Use}. At the same time, overt naming is highly informative because it preserves the natural link between object recognition and a spoken word response. A successful semantic-category decoding analysis in this setting can therefore test whether neural activity during naming contains information aligned with the semantic structure of the named objects.

Multivariate analyses have provided evidence that semantic information can be recovered from non-invasive electrophysiological recordings. Chan et al. decoded both semantic category and individual word information from combined EEG and MEG recordings, suggesting that word and category representations are distributed across space and time \cite{Chan2011Decoding}. More recently, Wilmskoetter et al. showed that prearticulatory scalp EEG can predict coarse semantic categories of naming responses in healthy speakers, with informative activity occurring before overt speech \cite{Wilmskoetter2023Semantic}. Related large-scale work on non-invasive language decoding has further shown the promise of modern learning methods, while emphasizing that robust word-level decoding remains challenging, especially when generalization is required across participants, devices, tasks, or unseen words \cite{Defossez2023Decoding,DAscoli2024Decoding}. These studies motivate approaches that combine cognitively meaningful semantic targets with compact neural representations.

A complementary development comes from representation learning. Text embeddings provide continuous word representations that capture semantic similarity, and neural foundation models increasingly provide compact EEG representations learned from large-scale data. In language neuroscience, distributional word vectors have been used to model semantic information in brain responses, supporting the idea that text-derived representational spaces can serve as useful semantic models \cite{Sassenhagen2020Traces}. In EEG analysis, large-scale pre-training approaches aim to reduce reliance on task-specific feature engineering by learning reusable neural representations \cite{Jiang2025NeuroLM}. These developments provide a practical way to define semantic category targets from picture labels and to test whether EEG activity during naming contains information aligned with those targets.

Here, we test whether semantic-category information is recoverable from high-density EEG during overt picture naming. We derive a compact semantic target space from the French picture labels using multilingual text embeddings, then decode those categories from channel-wise EEG embeddings extracted from early, naming-related, and combined temporal windows. This design links computationally defined semantic structure to task-timed neural activity, allowing us to assess whether category-level information is present across the temporal unfolding of spoken picture naming.

\section{Methods}

\subsection{Participants and picture-naming task}

The dataset was previously collected in the context of a picture-naming EEG study \cite{Volpert2023Characterization}. The analyzed dataset comprised sixteen native French-speaking male participants. All participants had normal or corrected-to-normal vision and hearing, and were right-handed according to the Edinburgh Handedness Inventory \cite{Oldfield1971assessment}. Exclusion criteria included any history of neurological or psychiatric disorders, drug addiction, or head trauma.

Participants performed an overt picture-naming task using black-and-white line drawings selected from the standardized Snodgrass and Vanderwart picture corpus \cite{Snodgrass1980standardized}. This corpus provides normative measures for name agreement, image agreement, familiarity, and visual complexity, making it suitable for controlled studies of visual object recognition and naming. Pictures were presented on a screen and participants were instructed to name each picture aloud. Speech responses were recorded and synchronized with the EEG signal.

\begin{figure}[!h]
\centering
    \includegraphics[width=\columnwidth]{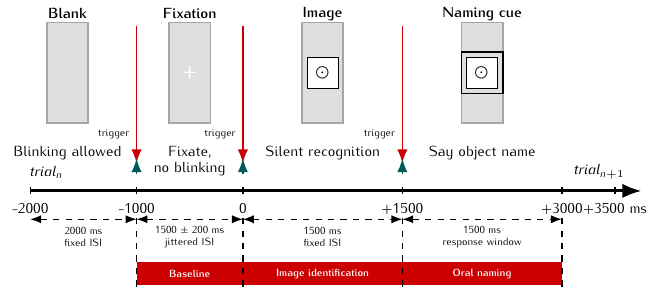}
    \caption{Single-trial structure of the picture-naming task. Participants could blink during the blank interval, fixated the cross without blinking, silently recognized the displayed object, and pronounced its name when the black frame appeared. Triggers were sent at fixation onset, image onset, and response-cue onset.}
    \label{fig:picture-naming-task}
\end{figure}

Each trial followed the structure illustrated in Fig.~\ref{fig:picture-naming-task}. Participants could blink during the initial blank interval, then fixated a central cross without blinking. After picture onset, they silently recognized the displayed object and prepared the corresponding name. They pronounced the name aloud only when the black frame appeared, which served as the response cue. Event triggers were sent at fixation onset, image onset, and response-cue onset, allowing the EEG signal to be aligned with the main task events.

Each participant completed 270 trials. Among these, 70 trials used a repeated control picture corresponding to a dog image, whereas 200 trials used distinct pictures associated with different target names. In the present study, the 200 distinct naming trials were used for semantic-category decoding. The repeated control trials were retained only for a preliminary control-versus-task separability check, which is reported in the Supplementary Material because this contrast confounds repetition, visual identity, lexical identity, and task structure.

\subsection{EEG acquisition and preprocessing}

EEG was recorded from 99 channels, comprising 96 EEG and 3 EOG/facial channels, at a sampling frequency of 1 kHz. The scalp distribution of the analyzed EEG channels is shown in Supplementary Fig.~\ref{fig:channel_locations}. The raw EEG signal was epoched around visual stimulus onset. In the original preprocessing pipeline, epochs had a duration of 5.5 s, spanning 2 s before and 3.5 s after picture onset. Epochs were baseline corrected using the interval from $-1$ s to 0 s relative to stimulus onset. Bad epochs containing artifacts such as eye blinks, eye movements, or head movements were removed, and the remaining epochs were band-pass filtered between 0.1 and 40 Hz. The final dataset is accessible from the public platform OpenNeuro \cite{Volpert2023Forced}.

For the analyses reported here, we focused on the 96 EEG channels and excluded non-EEG auxiliary channels from semantic decoding. The semantic analyses were restricted to the 200 unique picture-naming trials, since the 70 repeated control trials corresponded to a single repeated item and therefore did not define a multiclass semantic target.

\subsection{Semantic target construction}\label{sec:semantic_targets}

The semantic decoding task requires a target space that is both cognitively meaningful and statistically usable. A direct item-level classification of the 200 picture names would be underpowered, since each image appears only once per participant and each label would have very few observations. Conversely, manually assigning broad categories would introduce strong experimenter assumptions and could obscure the graded semantic relationships among items. We therefore constructed semantic targets from the lexical labels themselves, using text embeddings to derive a data-driven organization of the 200 picture names.

This choice follows a broader representational approach in cognitive neuroscience, where computational semantic spaces are used to relate linguistic meaning to neural activity. Distributional word vectors have been shown to capture semantic similarity and to model brain responses to words, supporting their use as external semantic representations \cite{Mikolov2013Efficient,Sassenhagen2020Traces}. Related work using representational similarity analysis has also emphasized that neural and computational representations can be compared through their internal similarity structure rather than through one-to-one label matching \cite{Kriegeskorte2008Representational,He2022Neural}. In the present study, text embeddings were not used as continuous regression targets, but as a basis for defining semantic classes for EEG classification.

Each of the 200 picture names was encoded with Qwen3-Embedding-0.6B, a multilingual text-embedding model designed for dense semantic representation and retrieval tasks \cite{Zhang2025Qwen3}. The use of a multilingual embedding model is appropriate here because the picture labels were French nouns, while the intended category structure should reflect semantic similarity rather than language-specific surface form. The resulting vectors provided a continuous semantic space in which semantically related labels were expected to lie closer together.

We then applied hierarchical agglomerative clustering with Ward's linkage to the text-embedding vectors. Ward's method merges clusters so as to minimize the increase in within-cluster variance, making it suitable for obtaining compact groups in a continuous embedding space. The dendrogram was used as an exploratory representation of the semantic structure among the 200 labels. Rather than treating the clustering output as a fully automatic ontology, we used it to guide the construction of a compact multiclass target space suitable for EEG decoding.

Fine-grained clusters were subsequently merged into nine interpretable semantic categories under constraints on the number of groups and their minimum size. The final categories were: \textit{DevicesVehicles}, \textit{FoodNature}, \textit{HouseholdObjects}, \textit{HumanRelated}, \textit{LandAnimals}, \textit{OtherAnimals}, \textit{PlacesSymbolsReading}, \textit{ToolsInstrumentsMedical}, and \textit{ToysSports}. This granularity level was chosen as a compromise between semantic specificity and class balance. It preserves distinctions that are relevant for picture naming, such as animals, tools, food, and human-related items, while avoiding a sparse item-level decoding problem. A two-dimensional visualization of the text-embedding space is provided in Supplementary Fig.~\ref{fig:words_clust_emb}.

\subsection{EEG representation and semantic decoding}

Recent EEG representation-learning models aim to reduce the dependence on handcrafted features by learning reusable neural representations from large-scale recordings. Early approaches such as BENDR used convolutional encoders and transformers with self-supervised objectives to learn from large amounts of unlabeled EEG data \cite{Kostas2021BENDR}. More recent EEG foundation models, including LaBraM and NeuroLM, extend this idea through large-scale pretraining, tokenization, and transformer-based modeling of EEG sequences \cite{Jiang2024LaBraM,Jiang2025NeuroLM}. These models are motivated by a common limitation of supervised EEG decoding: task-specific models often require substantial labeled data and can be sensitive to differences in montage, subjects, recording conditions, and downstream tasks.

In this work, we used SingLEM, a Single-Channel Large EEG Model, as a fixed EEG feature extractor \cite{Sukhbaatar2025SingLEM}. SingLEM is designed to learn general-purpose representations from individual EEG channels, rather than requiring a fixed multi-channel montage. This property is well suited to the present high-density picture-naming dataset, because each channel can be embedded independently while preserving channel-specific information. SingLEM combines local temporal feature extraction with hierarchical transformer modeling and was pretrained on a large heterogeneous EEG corpus, making it appropriate for downstream decoding with limited task-specific supervision.

For each trial and EEG channel, we extracted SingLEM representations from two task-relevant temporal windows. The \textit{EARLY} representation corresponds to the post-stimulus interval associated with visual and early lexical-semantic processing, whereas the \textit{NAMING} representation corresponds to the later interval associated with name preparation and overt response. Each window produced a 16-dimensional embedding per channel. Concatenating both windows yielded a 32-dimensional \textit{EARLY+NAMING} representation.

This channel-wise construction transformed each picture-naming trial into a set of channel-level samples. Thus, a single trial contributed one embedding per EEG channel for each temporal representation. This organization preserves spatial information at the channel level while keeping the feature dimensionality compact. The resulting feature sets were used to test whether semantic category information was present in the early window, the naming window, or their combination.

Semantic decoding was evaluated over the nine text-embedding-derived categories defined above. The repeated control trials were not part of this multiclass semantic target space. We considered three decoding tasks: \textit{EARLY}, using only SingLEM early-window features; \textit{NAMING}, using only SingLEM naming-window features; and \textit{EARLY+NAMING}, using the concatenation of both feature sets.

For each task, features were standardized within each cross-validation fold and classified using a $k$-nearest-neighbor classifier with $k=5$, distance weighting, and Euclidean distance. This classifier provides a direct test of neighborhood structure in the embedding space: high performance indicates that channel-level embeddings from the same semantic category tend to occupy nearby regions after standardization. The implementation also included a stratified dummy baseline for reference, while the reported semantic-decoding results focus on the KNN model.

Performance was estimated using stratified five-fold cross-validation at the channel-sample level. Class labels were encoded from the semantic category assignments, and folds were stratified to preserve the distribution of the nine categories across train and test partitions. The main metrics were balanced accuracy and Macro-F1, which summarize multiclass performance while giving equal weight to semantic categories. Because cross-validation was performed at the channel-sample level, the reported values quantify within-dataset semantic separability of channel-wise EEG embeddings rather than generalization to unseen participants, unseen picture items, or unseen trials.

\subsection{Control-versus-task separability check}

Before semantic category decoding, we evaluated whether compact EEG features could separate the repeated control item from the other picture-naming trials. Compact ERP and bandpower features were extracted and classified with a support-vector machine using stratified five-fold cross-validation. This analysis achieved an AUC of $0.793$ and is reported in Supplementary Fig.~\ref{fig:base_results}. Because the control condition is a repeated single item, this contrast should be interpreted as a quality-control and separability check rather than evidence for semantic decoding.

\section{Results}

\subsection{Semantic decoding across temporal windows}

Table~\ref{tab:semantic_results} reports nine-class semantic decoding performance obtained from the three SingLEM feature sets. All feature sets yielded performance above the chance level expected for nine balanced classes, i.e., approximately $1/9 = 0.111$. The \textit{EARLY} representation reached a balanced accuracy of $0.562 \pm 0.002$ and a Macro-F1 score of $0.566 \pm 0.002$. The \textit{NAMING} representation improved performance to $0.610 \pm 0.003$ balanced accuracy and $0.613 \pm 0.003$ Macro-F1. The best performance was obtained with the combined \textit{EARLY+NAMING} representation, which reached $0.781 \pm 0.002$ balanced accuracy and $0.784 \pm 0.002$ Macro-F1.

\begin{table}[!htb]
\setlength{\tabcolsep}{4pt}
\centering
    \caption{Semantic-category decoding during picture naming. Panel A reports global nine-class decoding metrics from the three temporal feature sets. Values are mean $\pm$ standard deviation across five stratified folds. Panel B reports class-level F1 scores pooled across the five folds.}
    \label{tab:semantic_results}
    \begin{tabular}{lccc}
    \toprule
    \multicolumn{4}{l}{\textbf{Panel A. Global nine-class decoding performance}} \\
    \midrule
    Feature set & Dim. & Balanced acc. & Macro-F1 \\
    \midrule
    \textit{EARLY} & 16 & 0.562 $\pm$ 0.002 & 0.566 $\pm$ 0.002 \\
    \textit{NAMING} & 16 & 0.610 $\pm$ 0.003 & 0.613 $\pm$ 0.003 \\
    \textit{EARLY+NAMING} & 32 & 0.781 $\pm$ 0.002 & 0.784 $\pm$ 0.002 \\
    \midrule
    \multicolumn{4}{l}{\textbf{Panel B. Class-level F1 scores}} \\
    \midrule
    Class & \textit{EARLY} & \textit{NAMING} & \textit{EARLY+NAMING} \\
    \midrule
    DevicesVehicles & 0.567 & 0.608 & 0.780 \\
    FoodNature & 0.555 & 0.616 & 0.782 \\
    HouseholdObjects & 0.567 & 0.606 & 0.774 \\
    HumanRelated & 0.578 & 0.610 & 0.786 \\
    LandAnimals & 0.569 & 0.617 & 0.787 \\
    OtherAnimals & 0.565 & 0.621 & 0.785 \\
    PlacesSymbolsReading & 0.551 & 0.604 & 0.780 \\
    ToolsInstrumentsMedical & 0.591 & 0.637 & 0.798 \\
    ToysSports & 0.551 & 0.596 & 0.786 \\
    \bottomrule
    \end{tabular}
\end{table}

The comparison between temporal feature sets shows a clear progression. The \textit{EARLY} window already contains discriminative information about the semantic category of the named object, consistent with the presence of visual and early lexical-semantic processing after picture onset. The \textit{NAMING} window performs better than \textit{EARLY}, suggesting that semantic-category structure becomes more separable during the later phase associated with name preparation and response production.

The strongest effect was observed when the two temporal windows were concatenated. The gain from \textit{NAMING} to \textit{EARLY+NAMING} was larger than the gain from \textit{EARLY} to \textit{NAMING}, indicating that the two windows provide complementary information rather than a redundant estimate of the same signal. This pattern suggests that category-relevant information is distributed across the temporal evolution of picture naming rather than confined to a single processing window.

Class-level F1 scores showed the same temporal pattern as the global metrics. All categories improved from \textit{EARLY} to \textit{NAMING}, and all reached their highest F1 score with the combined \textit{EARLY+NAMING} representation. The gain from \textit{EARLY} to \textit{EARLY+NAMING} was consistent across categories, with final F1 scores ranging from $0.774$ for \textit{HouseholdObjects} to $0.798$ for \textit{ToolsInstrumentsMedical}. This indicates that the improvement obtained by combining temporal windows was not driven by a single semantic category, but reflected a broad increase in class-level separability.

\subsection{Channel-level distribution of semantic decoding}

\begin{figure}[!ht]
    \centering
    \includegraphics[width=\columnwidth]{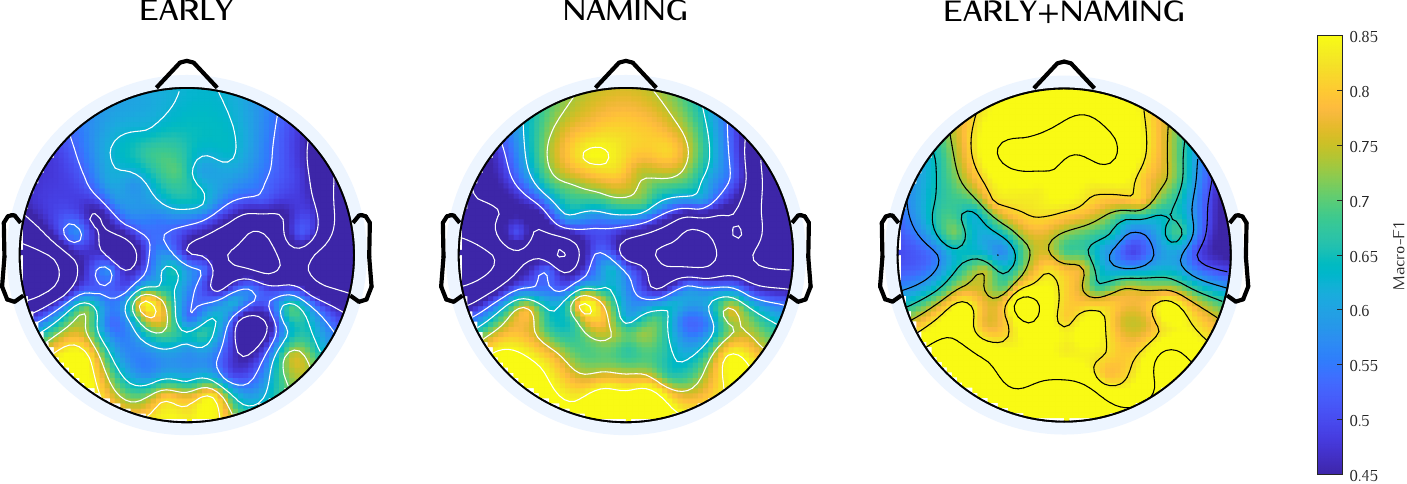}
    \caption{Channel-level semantic decoding topography. Scalp maps show Macro-F1 scores for the \textit{EARLY}, \textit{NAMING}, and \textit{EARLY+NAMING} SingLEM representations. All panels use the same color scale.}
    \label{fig:channel_f1_topoplots}
\end{figure}

Fig.~\ref{fig:channel_f1_topoplots} summarizes the spatial distribution of channel-level Macro-F1 scores for the three feature sets. The \textit{EARLY} representation shows moderate and spatially heterogeneous semantic separability, with higher values mainly over posterior and inferior channels and lower values over lateral-central channels. The \textit{NAMING} representation increases decoding performance, with stronger values over frontal and posterior channels. The \textit{EARLY+NAMING} representation yields the highest and most spatially widespread F1 scores, indicating that the improvement obtained by combining temporal windows is not restricted to a single channel cluster.

This channel-level pattern is consistent with the temporal structure of picture naming, in which visual object processing, semantic access, lexical selection, phonological encoding, and articulatory preparation unfold over successive processing stages \cite{Indefrey2004spatial,Indefrey2011spatial}. The stronger performance in the naming-related and combined representations also agrees with prior work showing that semantic categories of naming responses can be predicted from pre-articulatory scalp EEG activity \cite{Wilmskoetter2023Semantic}. Importantly, these maps should be interpreted as channel-level decoding topographies rather than anatomical source maps. Scalp EEG topographies are affected by volume conduction, reference choice, and montage geometry, and channel-space analyses alone do not justify precise cortical localization claims \cite{Lai2018Comparison,Liu2023Comparison}.

\section{Discussion}

This study tested whether semantic-category information is recoverable from high-density EEG during overt picture naming. The results show a consistent pattern: decoding was above chance for all temporal feature sets, improved from the \textit{EARLY} window to the \textit{NAMING} window, and reached its strongest performance when the two windows were combined. The best model achieved a balanced accuracy of $0.781$ and a Macro-F1 score of $0.784$ across nine semantic categories, indicating strong semantic separability in the channel-level representation space.

The main contribution is evidence that category-level semantic structure is reflected in EEG activity during spoken picture naming. The improvement from \textit{EARLY} to \textit{NAMING} and from \textit{NAMING} to \textit{EARLY+NAMING} suggests that the relevant information is distributed across the temporal progression of the task. Early activity may reflect visual object processing and the beginning of conceptual-semantic access, whereas later naming-related activity may include name preparation and response-related processes. The combined-window result indicates that these stages carry complementary category-relevant information.

The embedding-based pipeline provides a practical way to test this language-neuroscience question. On the target side, multilingual text embeddings provided a data-driven way to organize French picture labels into interpretable semantic categories. Compared with conventional word-vector approaches, recent embedding models are better suited to multilingual semantic similarity and clustering tasks \cite{Zhang2025Qwen3,Muennighoff2023MTEB}. On the neural side, SingLEM \cite{Sukhbaatar2025SingLEM} provided compact channel-wise EEG representations learned from large heterogeneous EEG datasets, reducing dependence on handcrafted features and fixed multi-channel architectures. In the present context, these methods are best understood as tools for asking whether semantic structure can be detected in EEG during a language-production task.

The present results are consistent with prior work showing that semantic categories of naming responses can be predicted from prearticulatory scalp EEG \cite{Wilmskoetter2023Semantic}. The present study extends this line of work by combining text-derived semantic targets with pre-trained single-channel EEG embeddings and by comparing early, naming-related, and combined temporal representations. The strong improvement in the combined representation supports the view that semantic-category information in overt naming is temporally distributed, rather than being expressed only in a single isolated interval.

Several limitations remain. First, because semantic categories are defined over picture labels, decoding may reflect a mixture of visual object properties, conceptual-semantic structure, lexical access, and response preparation. The present design therefore supports category-level decoding during picture naming, but does not isolate a process-pure semantic stage. Second, the dataset is modest in the number of participants and item repetitions, even though the number of channel-level samples is large. This is a common constraint in EEG studies, where signals are noisy, non-stationary, and strongly variable across participants. Inter-subject and intra-subject variability are known to limit the generalization ability of EEG decoding models \cite{Huang2023Discrepancy,Saha2020Intra}. Third, the present cross-validation estimates semantic separability within the observed channel-level representation space. A leave-one-subject-out protocol would address a different and harder question: generalization to entirely unseen participants. Future work should also test trial-grouped and item-held-out decoding to separate within-dataset semantic separability from generalization to new trials, new items, and new speakers.

In conclusion, this study provides evidence that semantic-category information is decodable from EEG during overt picture naming. By combining text-embedding-derived semantic targets with pre-trained single-channel EEG embeddings, we obtained strong nine-class semantic separability, with the best results achieved by integrating early and naming-related temporal information. These findings support the use of modern neural decoding methods as tools for investigating lexical-semantic processing in spoken language production.

\bibliographystyle{ieeetr} 
\bibliography{bib}

\newpage
\section*{Supplementary Material}
\setcounter{figure}{0}
\renewcommand{\thefigure}{S\arabic{figure}}
\renewcommand{\theHfigure}{S\arabic{figure}}

\begin{figure}[!ht]
    \centering
    \includegraphics[width=0.50\columnwidth]{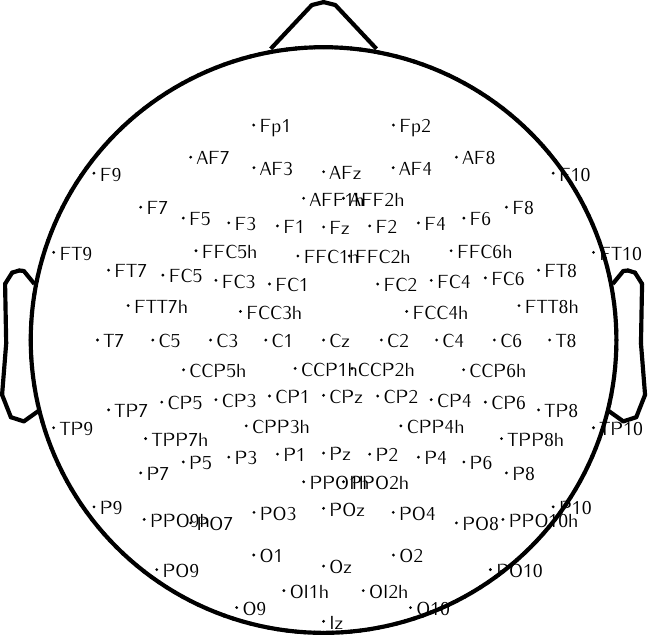}
    \caption{EEG channel layout used for analysis.}
    \label{fig:channel_locations}
\end{figure}

\begin{figure}[!ht]
    \centering
    \includegraphics[width=0.8\columnwidth]{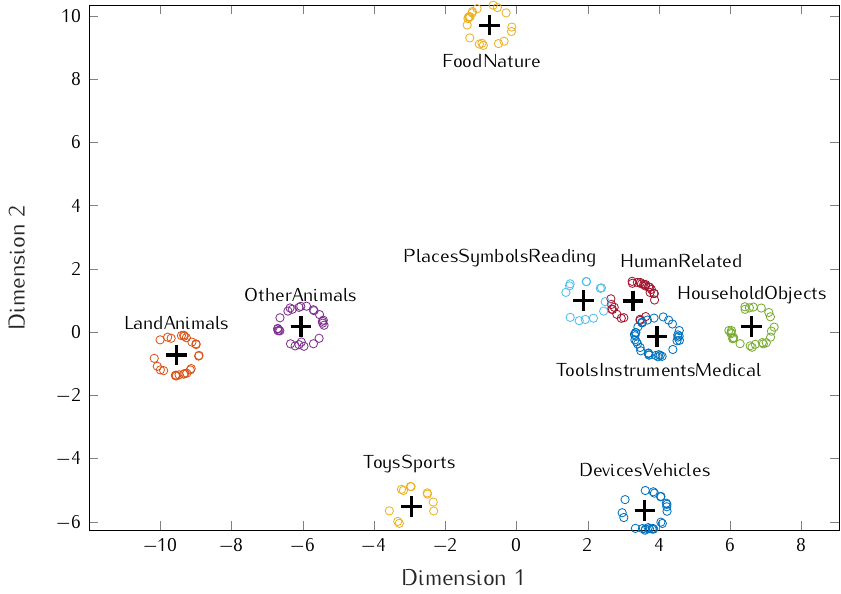}
    \caption{Semantic map of the 201 picture labels. Colored points show labels in a 2D t-SNE projection of text embeddings; black crosses indicate category centroids. The control item is shown once but excluded from semantic classification.}
    \label{fig:words_clust_emb}
\end{figure}

\begin{figure}[!ht]
\centering
    \includegraphics[width=\columnwidth]{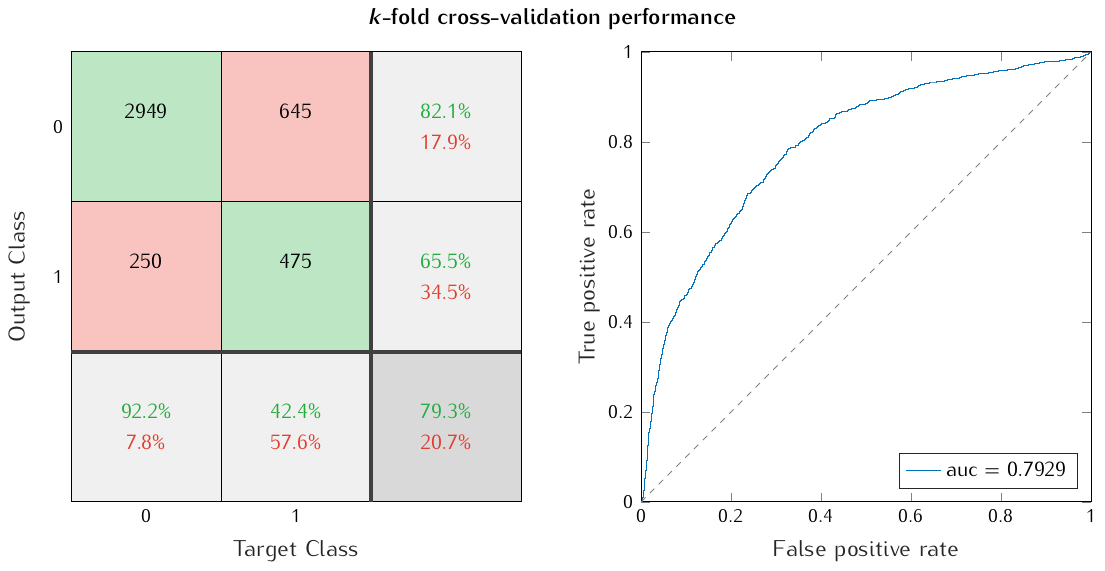}
    \caption{Control-versus-task classification performance. ROC curves for the SVM classifier evaluated with stratified five-fold cross-validation. This contrast was used only as a separability check and was not interpreted as semantic decoding.}
    \label{fig:base_results}
\end{figure}
\vfill

\end{document}